\documentclass[12pt]{article}

\usepackage{lmodern}
\usepackage{graphicx}             
\DeclareGraphicsExtensions{.pdf}  
\usepackage{tikz}                 
\usetikzlibrary{shapes,arrows}
\usepackage{subfig}            
\usepackage{amsmath}              
\usepackage{amssymb}              
\usepackage{amsthm}               
\usepackage[squaren,mediumqspace]{SIunits}     
\usepackage{multirow}             
\usepackage{tabularx}             
\usepackage{slashed}              
\usepackage{booktabs}             
\usepackage{import}               
\usepackage{xcolor}    
\usepackage{bigdelim}             
\usepackage{bbm}                  
\usepackage{feynmp}               
\usepackage{wasysym}              
\usepackage{cite}                 
\usepackage{hyphenat}             
\usepackage{datetime}
\usepackage[utf8]{inputenc}
\usepackage{ulem}

\newcommand{\htGWb}[1]{{\color{black} #1}}

\newcommand{\gsim}
{\;\raisebox{-.3em}{$\stackrel{\displaystyle >}{\sim}$}\;}

\newcommand{\fmfvcenterwl}[1]{\vcenter{\hbox{\qquad \fmfreuse{#1} \qquad}}}

\evensidemargin \oddsidemargin
\begin{document}

\thispagestyle{empty}

\def\thefootnote{\fnsymbol{footnote}}

\begin{flushright}
{DESY 20-083} \\
\end{flushright}

\vspace{0.5cm}

\begin{center}
{\large\sc {\bf Prospects for direct searches for light Higgs bosons\\[.7em]
at the ILC with 250~GeV}}

\vspace{0.5cm}

\vspace{1cm}

{\sc
P.~Drechsel$^{1}$%
\footnote{former address}%
, G.~Moortgat-Pick$^{1,2}$%
\footnote{email: gudrid.moortgat-pick@desy.de}%
,~and G.~Weiglein$^{1}$%
\footnote{email: georg.weiglein@desy.de}
}

\vspace*{.7cm}

{\sl
$^1$DESY, Notkestra\ss e 85, D--22607 Hamburg, Germany

\vspace*{0.1cm}

$^2$II.  Institut f\"ur  Theoretische  Physik, Universit\"at  Hamburg,\\
Luruper Chaussee 149, 22761 Hamburg, Germany

}

\end{center}

\vspace*{0.1cm}

\begin{abstract}
The particle discovered in the Higgs boson searches at the LHC with a mass of
about 125~GeV is compatible within the present uncertainties with the 
Higgs boson predicted in the Standard Model (SM), but it could also 
be identified with one of the neutral Higgs bosons in a variety of
Beyond the SM (BSM) theories with an extended Higgs sector.
The possibility that an additional Higgs boson (or even more than one) 
could be lighter than the state that has been detected at 125~GeV occurs
generically in many BSM models and has some support from slight excesses that
were observed above the background expectations in Higgs searches at LEP and
at the LHC.
The couplings between additional Higgs fields and the electroweak
gauge bosons in BSM theories could be probed by model-independent Higgs
searches at lepton colliders. We present a generator-level extrapolation
of the limits obtained at LEP to the case of a future $e^+e^-$ collider, 
both for the search where the light Higgs boson decays 
into a pair of bottom quarks and for the decay-mode-independent search
utilising the recoil method. 
We find that at the ILC with a centre-of-mass energy of 250~GeV, an integrated
luminosity of 500~fb$^{-1}$ and polarised beams, the sensitivity to a light
Higgs boson with reduced couplings to gauge bosons is improved by more than
an order of magnitude compared to the LEP limits 
and goes much beyond the
projected indirect sensitivity of the HL-LHC with 3000~fb$^{-1}$ from the
rate measurements of the detected state at 125~GeV.
\end{abstract}

\def\thefootnote{\arabic{footnote}}
\setcounter{page}{0}
\setcounter{footnote}{0}

\newpage



\newsavebox{\SE}
\sbox{\SE}{
\begin{fmffile}{Feynman/proc}

\begin{fmfgraph*}(58,28)
  \fmfkeep{eeToHZ}
  \fmfleft{i1,i2}
  \fmfright{o1,o2}
  \fmf{plain,tension=1.}{i1,v1,i2}
  \fmf{wiggly,tension=1.,label=\scalebox{0.8}{$Z$}}{v1,v2}
  \fmf{wiggly,tension=1.}{v2,o1}
  \fmf{dashes,tension=1.}{v2,o2}
  \fmflabel{\scalebox{0.8}{$e$}}{i1}
  \fmflabel{\scalebox{0.8}{$e$}}{i2}
  \fmflabel{\scalebox{0.8}{$Z$}}{o1}
  \fmflabel{\scalebox{0.8}{$H$}}{o2}
\end{fmfgraph*}

\begin{fmfgraph*}(58,28)
  \fmfkeep{eeToHPhiZ}
  \fmfleft{i1,i2}
  \fmfright{o1,o2}
  \fmf{plain,tension=1.}{i1,v1,i2}
  \fmf{wiggly,tension=1.,label=\scalebox{0.8}{$Z$}}{v1,v2}
  \fmf{wiggly,tension=1.}{v2,o1}
  \fmf{dashes,tension=1.}{v2,o2}
  \fmflabel{\scalebox{0.8}{$e$}}{i1}
  \fmflabel{\scalebox{0.8}{$e$}}{i2}
  \fmflabel{\scalebox{0.8}{$Z$}}{o1}
  \fmflabel{\scalebox{0.8}{$H/\phi$}}{o2}
\end{fmfgraph*}

\begin{fmfgraph*}(58,28)
  \fmfkeep{eeToPhiZ}
  \fmfleft{i1,i2}
  \fmfright{o1,o2}
  \fmf{plain,tension=1.}{i1,v1,i2}
  \fmf{wiggly,tension=1.,label=\scalebox{0.8}{$Z$}}{v1,v2}
  \fmf{wiggly,tension=1.}{v2,o1}
  \fmf{dashes,tension=1.}{v2,o2}
  \fmflabel{\scalebox{0.8}{$e$}}{i1}
  \fmflabel{\scalebox{0.8}{$e$}}{i2}
  \fmflabel{\scalebox{0.8}{$Z$}}{o1}
  \fmflabel{\scalebox{0.8}{$\phi$}}{o2}
\end{fmfgraph*}

\end{fmffile}
}

\section{Introduction \label{sec:1}}

The properties of the Higgs boson that was discovered in 2012 are in
agreement with the predictions of the Standard Model (SM) 
within the current experimental accuracy, but they are also compatible with a
wide variety of extensions of or alternatives to the SM. Extended Higgs
sectors predict the existence of additional Higgs bosons, which could be
heavier but also lighter than the observed state at 125~GeV. The
coupling of the SM Higgs boson to the gauge bosons $W$ and $Z$ is such that
terms with a bad high-energy behaviour in longitudinal vector-boson
scattering exactly cancel with each other. As a consequence, 
in an extended Higgs sector comprising the SM-like state at 125~GeV
and additional Higgs bosons the couplings of the additional neutral Higgs
bosons to $W$ and $Z$, $g_{\phi_i VV}$,
are expected to be small. This implies that additional
heavy neutral Higgs bosons may not be detectable via the search channels 
$\phi_i \to ZZ, W^+W^-$ and that for an additional light Higgs boson (or more than
one) $g_{\phi VV}$ may be so small that such a light Higgs boson would have
escaped the limits from the Higgs searches at
LEP~\cite{Barate:2003sz,Schael:2006cr} (see also Ref.~\cite{Abbiendi:2002qp})
and the Tevatron~\cite{Aaltonen:2012if}.

The case of an extended Higgs sector containing a SM-like state that can be
identified with the observed Higgs signal and further Higgs bosons of which
at least one is lighter than 125~GeV can be realised, for instance, 
in a general two-Higgs-doublet model%
\footnote{See e.g.\ Refs.~\cite{Bechtle:2016kui,Bahl:2018zmf} for recent
discussions of the viability of such a scenario in the Minimal
Supersymmetric extension of the SM (MSSM).}
(2HDM),
and it occurs generically in extensions with a light singlet such as the
Next-to-Minimal Supersymmetric extension of the SM (NMSSM), see
e.g.~Refs.~\cite{Domingo:2015eea,Drechsel:2016jdg,Domingo:2018uim}, or the
N2HDM, a 2HDM with an additional real Higgs singlet,
see e.g.~Ref.~\cite{Biekotter:2019kde}.
It should be noted that extensions of the SM with a singlet-dominated state 
in the mass range around or just below 100~GeV are also of interest in view
of the observed local excesses around 96~GeV at the 2--3$\,\sigma$ level
in both the searches at LEP 
in the $e^+e^- \to Z\phi$, $\phi \to b \bar b$ 
channel~\cite{Barate:2003sz,Schael:2006cr}
and at CMS in the
$\phi \to \gamma\gamma$ searches~\cite{CMS:2017yta} (the CMS result
is compatible~\cite{Heinemeyer:2018wzl} with
the results of the corresponding searches at ATLAS~\cite{ATLAS:2018xad}).
Possible interpretations have been discussed e.g.\ in the 
NMSSM~\cite{Domingo:2018uim,Choi:2019yrv,Cao:2019ofo}, 
an inflation-inspired $\mu$NMSSM~\cite{Hollik:2018yek,Hollik:2020plc}
and the N2HDM~\cite{Biekotter:2019kde}.

Specifically, the neutral Higgs bosons $\phi_i$ of extended Higgs sectors consisting 
of any number of doublets and singlets fulfill the sum rule at lowest
order that the squared couplings to gauge bosons of all $\phi_i$ add up to
the squared coupling of the SM Higgs boson to gauge bosons,
\begin{equation}
\sum_i (g_{\phi_i VV})^2 = (g^{\rm SM}_{HVV})^2 ,\quad\mbox{\rm where }
\; V\in\{W,Z\}.
\label{eq:g2}
\end{equation}
Accordingly, this sum rule is valid for a wide class of models, including all
the examples of
Beyond the Standard Model (BSM) theories with an extended Higgs sector
mentioned above. The sum rule receives corrections at the loop level, but
these amount to effects that are typically at the per cent level.
This pattern of extended Higgs sectors implies on the one hand that the
couplings of the observed state at 125~GeV should be measured with the
highest possible precision in order to maximise the sensitivity for
establishing a deviation from the SM values, and on the other hand it provides a
strong motivation to search for additional Higgs bosons with couplings to
gauge bosons that are significantly suppressed compared to the case
of a SM-like Higgs boson with the
same mass. As far as the search for light additional Higgs bosons is
concerned, the mass region between about 60~GeV (since for $2 m_{\phi} >
125$~GeV the decay of the Higgs boson at 125~GeV, $h(125)$, into a pair of
the additional Higgs bosons, $h(125) \to \phi\phi$, is kinematically closed)
and 100~GeV appears to be particularly promising, as this mass range is
only mildly constrained by the existing limits from the Higgs searches at LEP
and the $\phi \to \gamma\gamma$ searches at the LHC (and as discussed above,
in both types of these searches an interesting excess above the background
expectation has been reported).

In the current paper we study the sensitivity of searches at 
the ILC in its first stage of $\sqrt{s}=250$~GeV centre-of-mass energy for light 
additional Higgs bosons with masses below the one of the observed signal at
125~GeV.
Due to the clean environment and consequently a very favourable
signal-to-background ratio for the prospective integrated luminosity, 
the low beamstahlung, the precise knowledge of
the beam energy and the availability of polarised beams, the ILC 
has a high physics potential in the direct search for 
such light additional Higgs bosons. 
We perform a generator-level extrapolation
of the limits obtained at LEP to the ILC case, both for the search 
for the $\phi \to b \bar b$ final state and for the
decay-mode independent recoil technique (the latter was pioneered at LEP 
by the OPAL collaboration~\cite{Abbiendi:2002qp} and forms the basis for 
the total cross-section measurement for $Zh(125)$ production at future
$e^+e^-$ colliders).

The two main production processes for a neutral Higgs boson at the ILC
are Higgs-strahlung ($e^+e^-\to (H/\phi) Z$), dominant at lower masses and
lower collider energies, and $WW$-fusion ($e^+e^-\to (H\phi) \nu \bar{\nu}$),
dominant at higher Higgs masses and higher collider energies.
In the current study, we therefore focus on the Higgs-strahlung process, for
which we investigate the $\phi \to b \bar b$ channel as well as the recoil
method, where only the leptonic $Z$-boson decays are utilised for the
reconstruction of the final state (in our study we have concentrated on the 
$Z\to \mu^+\mu^-$ decay). We first validate our approach with the LEP results and
then perform an extrapolation to the case of the ILC.
The present paper builds up on a preliminary study that was carried out in
Ref.~\cite{Drechsel:2018mgd}, see also Ref.~\cite{georg-mext}.

The paper is organised as follows. The adopted statistical method is
described in
Sect.~\ref{sec:2}. In Sect.~\ref{sec:3} the method is validated with the
results obtained at LEP, and in Sect.~\ref{sec:4} the sensitivity of
the ILC searches for light Higgs bosons is discussed. Our conclusions are given in
Sec.~\ref{sec:5}.

\section{Description of the applied methods \label{sec:2}}

We consider the ``Higgs-strahlung'' process
\begin{align}
  \quad\nonumber\\
  \fmfvcenterwl{eeToHPhiZ}\\
  \quad\nonumber
\end{align}
\noindent
where for our generator-level analysis we focus on the $Z$-boson decay into
muons, $Z \to \mu^+\mu^-$, and treat the cases where either the decay of the
scalar $H / \phi \to b \bar b$ is reconstructed,
\begin{equation}
e^+ + e^- \rightarrow Z +  (H/\phi) \rightarrow b  + \bar b +  \mu^- + \mu^+,
\label{eq:proc}
\end{equation}
or only the information from the leptonic $Z$-boson decay is used. For the
scalar either the SM Higgs boson $H$ is considered or a light Higgs boson
$\phi$ with a reduced coupling to gauge bosons compared to the case of a
SM-like Higgs boson with the same mass.

Concerning the statistical treatment we follow the prescription that was
outlined for $S_{95}$ in Refs.~\cite{Barate:2003sz, Bock:2004xz}, using
a simplified approach, see also~\cite{Bechtle:2004ne}. We have generated
event  samples corresponding to the  two hypotheses  ``all events  are
generated by the background only'' or  ``all events are generated by the
background plus a hypothetical signal''. \htGWb{This has been done 
by comparing the events for the full $e^+e^- \to 4 \; \mbox{fermion}$
process for the considered mass value of the assumed signal, taking into
account all interference contributions, with the events for the full $e^+e^-
\to 4 \; \mbox{fermion}$ process for the case where the mass value of the
assumed signal is beyond the kinematic reach of the collider.}
The quantity
\begin{align}
  \label{eq:DefS95}
  S_{95} = \frac{\hat{\sigma}}{\sigma_{\mathrm{ref}}} = \frac{\hat{n}}{n}
\end{align}
\noindent
gives  an  upper limit  $\hat{\sigma}$  on  a cross section  that  is
compatible  with   the  ``background  only''  hypothesis   at  the  95\%
confidence   level,   normalised   to  a   reference   cross   section
$\sigma_{\mathrm{ref}}$, or equivalently on the ratio of the
allowed signal
rate $\hat{n}$, normalised to the \htGWb{reference} signal rate $n$.
As reference  process  we use in Eq.~(\ref{eq:proc})  the case of the SM Higgs
boson at the considered mass value.
Accordingly, Eq.~(\ref{eq:DefS95})
can be  interpreted as  the
ratio  between the  squared couplings of  the scalars $\phi$ and $H$ to 
the $Z$-boson, 
\begin{align}
  S_{95} \mathrel{\hat=} \left|\frac{g_{\phi ZZ}^2}{g_{H ZZ}^2}\right| ,
  \label{eq:s95-g2}
\end{align}
for each mass value.

We have generated event samples for the process
\begin{equation}
e^+ + e^- \rightarrow b+ \bar b + \mu^- + \mu^+
\end{equation}
with the Monte Carlo  generator
\texttt{Whizard-2.4.1}~\cite{Kilian:2007gr,  Moretti:2001zz},  which we apply
for both signal processes and their reference values 
(with and without reconstruction of the $H / \phi \to b \bar b$ decay) as well
as for the corresponding backgrounds \htGWb{(a simple cut has been applied to
control backgrounds with photon radiation).}
Since we generate the full $e^+e^- \to 4 \mbox{ fermion}$ process, the
corresponding signal and background events are not generated
separately.  
The obtained events are classified by either the invariant mass of the quark
system or the recoil mass
of the lepton system  for this analysis.  
  The  
  signal rate in the $i$-th bin (we use a width of 1~GeV for the mass bins)
is determined by the difference of the event rates for the
  background plus  signal ($sb_i$) and the  background only hypotheses
  ($b_i$),  and  the  signal  rate   $s_i$  is  obtained  by  $sb_i  -
  b_i$.  Negative   rates  for  signal   events  are  cut,   $s_{i}=\max{\left(0,\ sb_i - b_i\right)}$.

We  assume that  the number  of potential
signal events $d_i$ in each of  the $i$ bins are distributed according
to  a Poisson  distribution with  the  expected values  $b_i$ for  the
``background only''  and $s_i  + b_i$ for  the ``background  plus signal''
hypotheses, respectively. The test statistic
\begin{equation}
  Q = \frac{L_{s+b}}{L_b}
  =  e^{-n} \prod\limits_{\text{bins } i}{
    \frac{\left(s_i + b_i\right)^{d_i}}{\left(b_i\right)^{d_i}}}
\label{eq:Q}
\end{equation}
orders the outcome of
test  experiments according  to  their ``signal  likeness''.  While  the
expectation values of the Poisson  distributions have to be determined
\textit{a priori}, the  numbers of potential signal  events $d_i$ have
to be determined experimentally.  For  this work we consider only
simulated  events,  and  thus  $d_i  = s_i$.   The  logarithm  of  the
test-statistic yields the weights for  the number of potential signal
events per bin as
\begin{align}
  \label{eq:Weight}
  w_i = \log{\left(1 + \frac{s_i}{b_i}\right)}.
\end{align}
\noindent
This definition requires the presence of at least one background event
per considered bin. In order to accommodate  this, we choose the luminosity for
the simulation large enough that at least one event can be
found  in each  bin,  i.e.  $b_i  \geq 1$,  and  scale the  luminosity
afterwards with a factor $c$ to  the desired value.  The weights $w_i$
are not affected by the scaling procedure.

With the  weight factors  of Eq.~\eqref{eq:Weight}, the  scaling factor
$S_{95}$  given  in  Eq.~\eqref{eq:DefS95}   can  be  expressed~\cite{Bock:2004xz}  
as
\begin{align}
  S_{95} = \frac{\hat{n}}{n}=\frac{K \cdot \sigma_{sb}}{\langle X \rangle_s}=\frac{1.96 \cdot \sigma_{sb}}{\langle X \rangle_s},
\end{align}
where $K$ denotes the number of standard deviations for the required
significance:  for $S_{95}$ the value of 
$K=1.96$ is used, corresponding to a confidence level (C.L.) of 5\% for the signal
hypothesis. The variance $\sigma_{sb}$ is given by $\sigma^2_{sb} =
\sum_{\text{bins } i}{w_i^2 \left( s_i + b_i \right)}$, and ${\langle X
\rangle_s}= \sum_{\text{bins } i}{w_i s_i}$.

\section{Validation of the methods with results from LEP I and LEP II \label{sec:3}}

In order to validate the described 
method we apply it in a first step to the search in the 
$\phi \to b \bar b$ channel and compare with the results
obtained by the LEP combination~\cite{Barate:2003sz,Schael:2006cr}. Furthermore, 
in a second step, we validate our approach for the case of the search via the
recoil method by comparing with the results obtained by the
OPAL collaboration in their
Higgs analysis employing the recoil method~\cite{Abbiendi:2002qp}.

We simulated the process  $e^+ + e^- \rightarrow b + \bar  b + \mu^+ +
\mu^-$    with    the    SM    implementation    of    \texttt{Whizard
  2.4.1}~\cite{Kilian:2007gr,  Moretti:2001zz}. This  process contains
for the signal events the Higgs-strahlung process where the produced Higgs
boson decays into $b \bar b$ and the $Z$ boson decays into $\mu^+\mu^-$. As
described above, for simplicity we restrict to the $b \bar b\mu^+\mu^-$ final
state, i.e.\ we do not perform a separate analysis for the 
$b \bar b e^+ e^-$ final state and for our analysis via the recoil method we
also do not simulate additional decay modes of the Higgs boson.
\htGWb{As explained above,}
our  ``background only''  and  ``background plus  signal'' hypotheses  are
generated  by   appropriately adapting  the  mass   of  the
implemented \htGWb{SM-like} Higgs  boson, i.e.\ for the
``background only''  hypothesis  the  mass  $m_\phi$ of  the  scalar  
is chosen beyond  the  kinematic
accessibility of the  LEP experiments. 

For LEP I we consider the integrated luminosity of the four experiments at 
the centre-of-mass  energy of  91.2~GeV.  For LEP  II we  use for the 
comparison the integrated luminosity  that the four LEP experiments recorded 
in a particular year and assign it to the highest energy that was achieved
during that year, with the exception of the last year of LEP running. For the
latter we use the energy of 206~GeV~\cite{Assmann:2002th}, where the bulk of
the luminosity was recorded during that year.  The
luminosity per experiment that we have used in our comparison is given for
the different LEP energy stages in
Tab.~\ref{tab:IntegratedLuminosityLEP}.
The number of events that we actually generated for our simulation is in fact 
400 times higher than the LEP luminosity in order to facilitate the
determinaton of the 
weight factors $w_i$, see Eq.~\eqref{eq:Weight},
and to reduce the statistical error. The results were then scaled down to the
appropriate luminosities as described above.

\begin{table}[h]
  \centering
  \caption{The integrated luminosities per experiment, 
$\int{\mathrm{d}t  \mathcal{L}}$, at the different stages of the LEP
centre-of-mass energies,  $\sqrt{s}$, that are used for the comparison with
the LEP results (see text).}
  \label{tab:IntegratedLuminosityLEP}
  \vspace{-.5cm}
  \begin{tabular}{lcccccc}
    \\\toprule
    $\sqrt{s}$/GeV & 91.2  & 172  & 184  & 189  & 202  & 206
    \\\midrule
    $\int{\mathrm{d}t \mathcal{L}}$/pb${}^{-1}$ & 208.44 & 24.7 & 73.4 & 199.7 & 253 & 233.4 
    \\\bottomrule
  \end{tabular}
\end{table}

In our generator-level analysis we do not take into account 
hadronisation effects of the $b$-quarks and we also do not simulate detector
effects.  This simplification leads to an over-optimistic estimate for
the signal  efficiencies and thus  for the observed signal  rates.  Since for
the searches making use of the $\phi \to b \bar b$ final state the signal
efficiencies have not explicitly been given
for the LEP experiments~\cite{Bechtle:2004ne}, 
we introduce a scale factor multiplying the luminosity that we use for the 
comparison with the LEP~1 results and we also allow for such a factor for the
comparison of the LEP~2 results. While our over-optimistic treatment of the
signal efficiencies implies the need to scale down our effective
luminosities, we also need to appropriately scale them up to account for the
fact that we have simulated only events for the $Z \to \mu^+\mu^-$ final
state, whereas the LEP analyses also incorporate the other decay modes of the
$Z$ boson. Both effects can be combined into a single scale factor. We
introduce such a factor, $c_{\rm bb}$, 
for the $\phi \to b \bar b$ analyses both at LEP~1 and LEP~2. The
corresponding factors for the recoil method analyses at 
LEP~1 and LEP~2 are denoted as $c_{\rm recoil}$. Specifically, we obtain the
effective luminosities $\mathcal{L}_{\mathrm{eff}}$ for our comparison as
\begin{align}
  \mathcal{L}_{\mathrm{eff}}^{\rm bb} = c_{\rm bb} \cdot \mathcal{L} ,
\qquad
  \mathcal{L}_{\mathrm{eff}}^{\rm recoil} = 
   c_{\rm recoil} \, \epsilon \cdot \mathcal{L},
\label{eq:Leff}
\end{align}
\noindent
where $\mathcal{L}$ is  the luminosity derived  from the values
given in Tab.~\ref{tab:IntegratedLuminosityLEP}.   The
scaling factors $c_{\rm bb},$ $c_{\rm recoil}$  depend on the parameters  of the experiment
and the analysis. We have determined those factors  from a comparison
with   the  LEP-combined results  for  $S_{95}$ 
in the case of the $\phi \to b \bar b$ analyses and with the OPAL results for
the case of the recoil method analyses. The four approximate
scaling factors, rounded to integer values, that we have
obtained in this way are listed in Tab.~\ref{tab:ScalingFactors}.

\begin{table}[h]
  \centering
  \caption{Luminosity scaling  factors $c_{\rm bb}$ and $c_{\rm recoil}$
derived via comparison with the $S_{95}$ results for the LEP~1
(91.2~GeV) and LEP~2 ($> 91.2$~GeV) energy stages from the LEP combination 
for the analysis 
using the $\phi \to b \bar b$ final state and from the OPAL results for
the recoil method analysis.}
  \label{tab:ScalingFactors}
  \vspace{-.5cm}
  \begin{tabular}{lcc}
    \\
    $\sqrt{s}$/GeV & 91.2 & $>91.2$
    \\\toprule
    $\phi \to b \bar b$ & $c_{\rm bb} = 12$ & $c_{\rm bb} = 4$
    \\\midrule
    recoil method & $c_{\rm recoil} = 4$ & $c_{\rm recoil} = 1$
    \\\bottomrule
  \end{tabular}
\end{table}

Obviously such a simple scaling factor can only roughly approximate 
an actual experimental analysis. For the case of the 
$\phi \to b \bar b$ analysis the scaling factor for incorporating besides
the $Z \to \mu^+\mu^-$ final state also the events for 
the other $Z$ decay modes would roughly correspond to 
an increase of the effective luminosity by a factor of 30 for an ideal 
detector. Assuming for simplicity a median signal efficiency  
of ~50\% for the signal rates leads to the conclusion that the coefficients 
$c_{\rm bb}$ in Tab.~\ref{tab:ScalingFactors} should not be
larger than $\approx 15$. The comparison with the listed results for $c_{\rm
bb}$ shows that the value that we obtained for the LEP~1 case (91.2~GeV) is
rather close to the expectation for this idealised case. 
\htGWb{For the LEP~2
case ($> 91.2$~GeV), on the other hand, the detector effects play a larger
role, and our generator-level extrapolation corresponds to a more optimistic
estimate in comparison with the analysis incorporating a realistic treatment
of the backgrounds and the experimental efficiencies}.
Tab.~\ref{tab:ScalingFactors} shows that for the recoil method the required
correction factors $c_{\rm recoil}$ 
are smaller. This is expected from the facts that the OPAL
analysis only made use of the 
decay modes $Z\to \mu^+\mu^-, e^+e^-$,  that for this analysis explicit signal 
efficiencies $\epsilon$ have been published, which we have taken into account
in Eq.~\eqref{eq:Leff}, and that generally the detector effects are expected 
to be less important for
the recoil analysis as it only relies on a leptonic final state.

\subsection{Comparison with LEP data using the $H$/$\phi$ decay mode} 

In order to compare our analysis for the search using the 
$H$/$\phi$ decay mode with the LEP data, we use the 
event samples for the process 
$e^+ + e^- \rightarrow b + \bar  b + \mu^+ + \mu^-$
and employ the information from the reconstruction of the 
$b \bar  b$ system for the identification of the samples for 
``background  only'', $b_i$,  and  ``background  plus
signal'',  $b_i +  s_i$. The information from the 
$\mu^+\mu^-$ system is only used for validating  that the signal events are
compatible with the production of a Higgs boson together with a
$Z$~boson.
A challenging region for this analysis is where the scalar mass 
$m_\phi$ is close to $M_Z$, since the presence of a large 
number  of background events close to the
  $Z$-boson mass $M_Z$ weakens the limit on  $S_{95}$; in this region we
  fit  the  expected background  events  to  emulate a  more  detailed
  analysis: we extrapolate  the expected background events  in an interval
    around $M_Z$ and fit these  points with a third-order  polynomial to obtain
    the number of events in the bins  around $M_Z$ as the value of the
    fit function  at the central mass  of each bin.

\begin{figure}
    \centering
    \includegraphics[width=.7\textwidth]{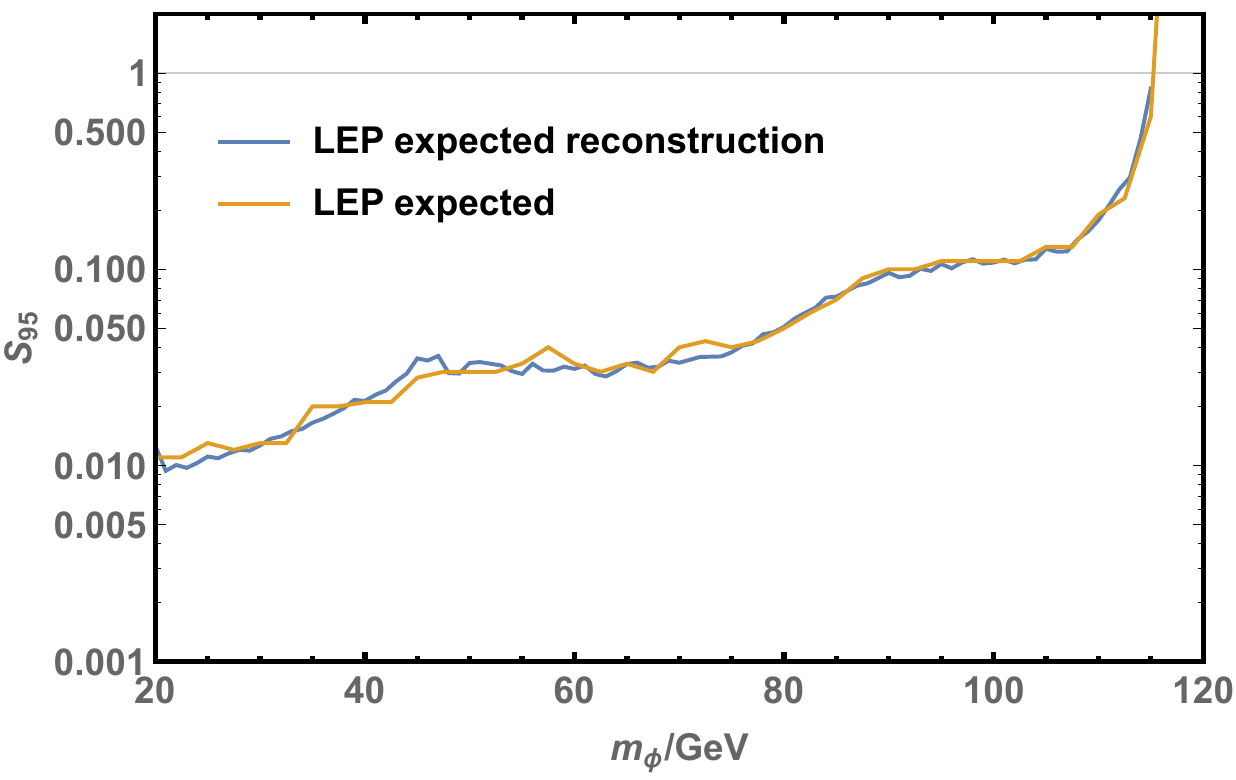}
\caption{The expected limit from LEP for the analyses using the $\phi \to b
\bar b$ decay information is compared with 
our method for approximately reconstructing the expected 
limit for $S_{95}$ 
from the process $e^+e^-\to Z (H/\phi) \to \mu^+\mu^- b \bar{b}$.
Our approximate result makes use of the two
scaling factors $c_{\rm bb}$ given in Tab.~\ref{tab:ScalingFactors}.}
\label{fig:LEPreconstruction-trad}
\end{figure}

The result using our method for approximately reconstructing the expected LEP
limit is shown in Fig.~\ref{fig:LEPreconstruction-trad} in
comparison with the expected limit that was published by the LEP
collaborations. As explained above, our method makes use of the 
two scaling factors $c_{\rm bb}$ given in
Tab.~\ref{tab:ScalingFactors}.
The comparison in Fig.~\ref{fig:LEPreconstruction-trad} shows that with those
two factors as input our method reproduces very well 
not only the normalisation of the 
expected LEP limit but also its shape as a function of 
$m_\phi$.

\subsection{Comparison with OPAL data using the recoil method}

The OPAL collaboration also used the recoil method for analysing the data,
i.e.\ exploiting the recoil of $H,\phi$ from the $Z$-boson and
analysing the $Z$ decay only. This method has the great advantage of being
completely independent of the $H,\phi$ decay modes. 

The OPAL analysis used the decay modes $Z\to \mu^+\mu^-$, $e^+e^-$~\cite{Abbiendi:2002qp}. 
We restrict our analysis to the decay mode $Z\to \mu^+\mu^-$ only, but
extrapolate in a second step the luminosity taken by OPAL to the full LEP
luminosity, see Tab.~\ref{tab:IntegratedLuminosityLEP}. We include only bins
close to $M_Z$ (interval [84~GeV,   98~GeV]), so that 
the  weighted mean  of  the
  central masses of the bins is in a small interval around $M_Z$
  \begin{align}
    \frac{\sum_{i=1}^{N}{d_i m_{i}}}{\sum_{j=1}^{N}{d_j}} \in [91.1~\text{GeV},\ 91.3~\text{GeV}]
  \end{align}
  \noindent
  with the number of  events $d_i$ in the $i$-th of  $N$ bins with the
  central mass $m_i$.

The events  from the $\mu$-lepton pairs  are ordered by
  their total energy $E_i$ into bins, and for each  bin we calculated
  the respective
  recoil mass $m^{(\text{rec})}_i$,
  \begin{align}
    m^{(\text{rec})}_i = \sqrt{s + M_Z^2 - 2 E_i \sqrt{s}},
  \end{align}
  \noindent
  to   obtain  the   event  rates   for  background,  $b_i   \equiv
  b_i(m_i^{(\text{rec})})$,      and      signal,     $s_i      \equiv
  s_i(m_i^{(\text{rec})}).$

Since the  OPAL collaboration published the signal  efficiency of their
analysis in
  Ref.~\cite{Abbiendi:2002qp},  we could make use of this information 
to estimate  a  mean  signal efficiency  of
  $\varepsilon   =   30\%$  for our approach and obtained   in this way 
the   effective   luminosity
  $\mathcal{L}_{\text{eff}}$ 
according to 
Eq.~\eqref{eq:Leff}.
The  obtained values for  the two scaling  factors 
$c_{\rm recoil}(\sqrt{s})$ are  given in
  Tab.~\ref{tab:ScalingFactors}.  

\begin{figure}
    \centering  
    \includegraphics[width=.7\textwidth]{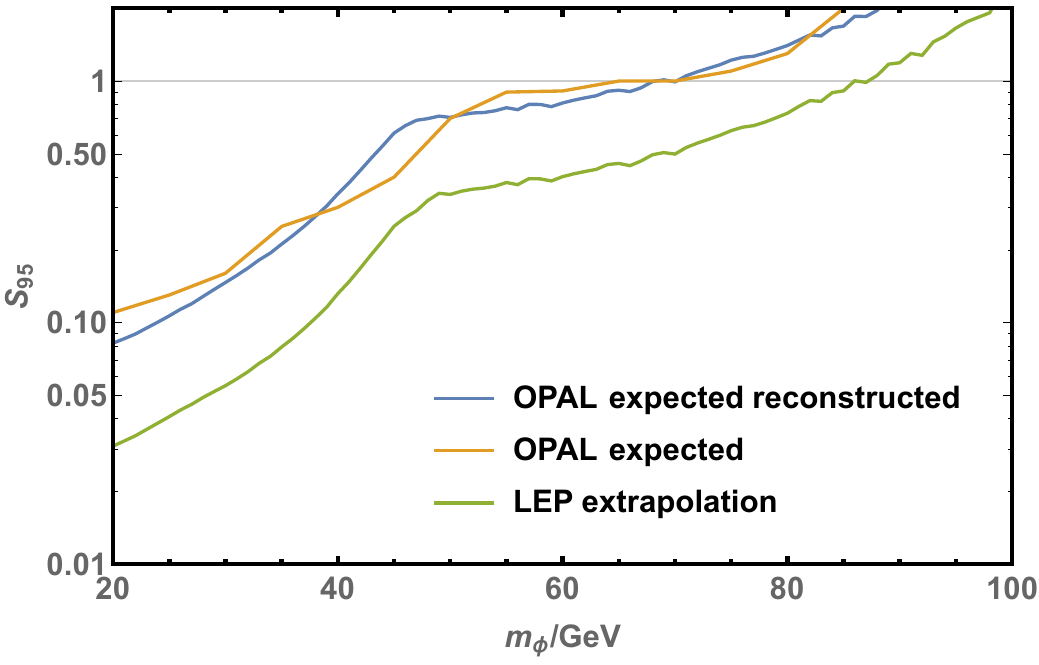}
\caption{The expected limit from OPAL based on the recoil method 
is compared with our method for approximately reconstructing the expected 
limit for $S_{95}$, 
which makes use of the two
scaling factors $c_{\rm recoil}$ given in Tab.~\ref{tab:ScalingFactors}.
Also shown is our result where the luminosity has been 
extrapolated  to  the  full  LEP luminosity.}
    \label{fig:LEPreconstruction-recoil}
\end{figure}

Our approximate reconstruction of the expected limit for 
$S_{95}$ from the recoil method analysis is compared with the expected 
limit published by the OPAL collaboration 
in Fig.\ref{fig:LEPreconstruction-recoil}, showing overall a good agreement. 
We furthermore display our result where 
the luminosity has been 
extrapolated  to  the  full  LEP luminosity.

\section{Discovery potential at the ILC for a light Higgs boson\label{sec:4}}

After having validated our method with the existing results from the analyses
at LEP data, we are now in a position to apply this method in order to derive
the expected limits for the ILC at $\sqrt{s}=250$~GeV for the two types of
analyses making use of the $\phi \to b \bar b$ decay information and
employing the recoil method. 
For our ILC analysis we use the two
scaling factors $c_{\rm bb}$ and $c_{\rm recoil}$ that we determined for LEP~2
as given in Tab.~\ref{tab:ScalingFactors}.
Concerning the ILC 
we assume a beam polarisation of $P_{e^-}=-80\%$ for the electron beam and
$P_{e^+}=+30\%$ for the positron beam, corresponding to the baseline
design~\cite{Adolphsen:2013kya}.  At the ILC with $\sqrt{s}=250$~GeV a total
luminosity of ${\cal L}=2000$~fb$^{-1}$ is expected to be collected within 15
years~\cite{Fujii:2017vwa}. For our study, however, we exploit only the
polarisation configuration $(-80\%,+30\%)$ and  assume the rather modest 
luminosity of 500~fb$^{-1}$.

  \begin{figure}[htb]
    \centering
  \includegraphics[width=.7\textwidth,
angle=-90]{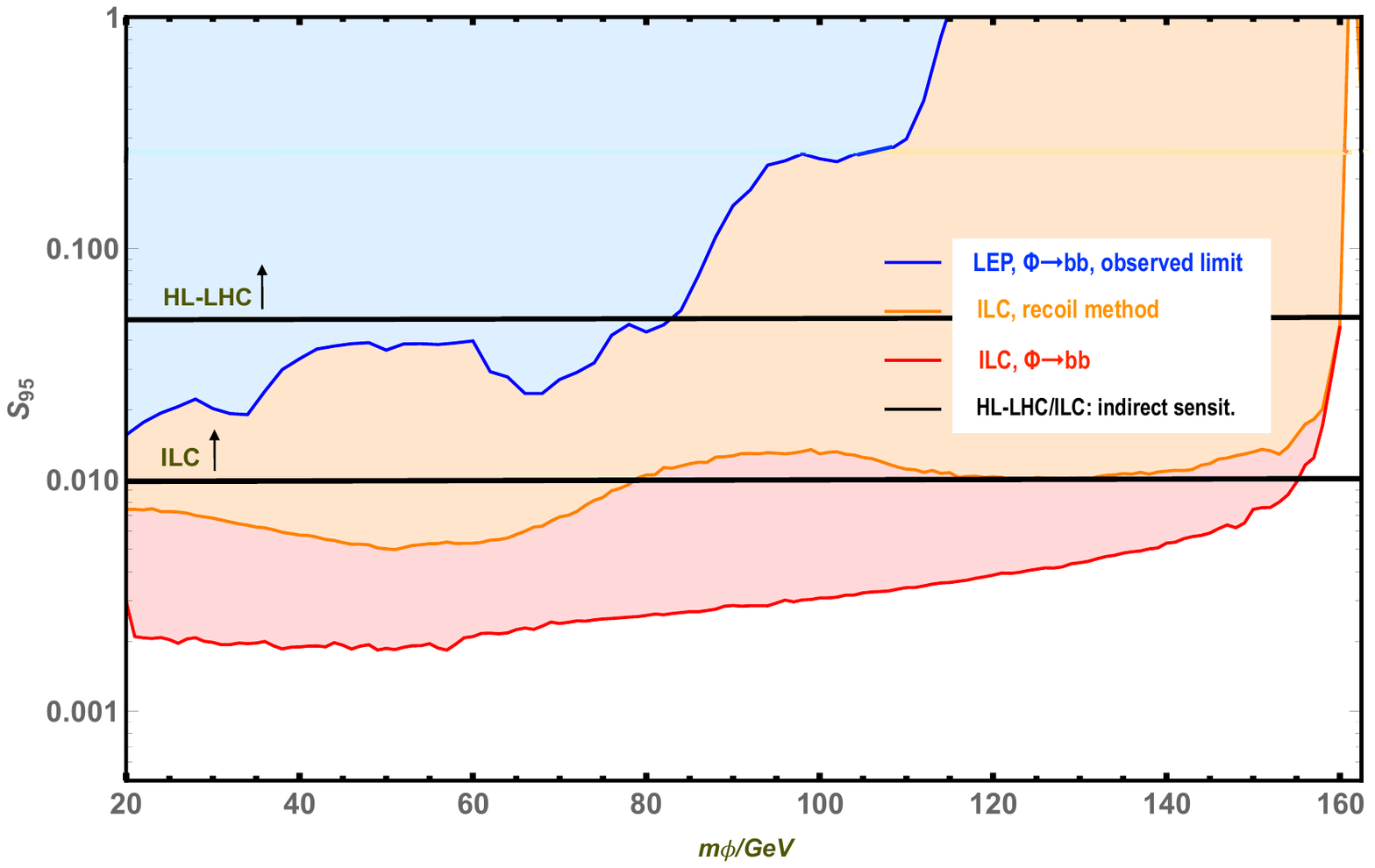}
    \caption{Projections for the expected limits on 
$S_{95}$ at the ILC~\cite{georg-mext} 
with $\sqrt{s}=250$~GeV, ${\cal L}=500$~fb$^{-1}$ and
polarised beams ($P_{e^-}=-80\%$ and $P_{e^+}=+30\%$) for the analysis using
the decay $\phi \to b \bar b$ and for the recoil method analysis, based only
on the recoil from the $Z$ boson that is reconstructed from the 
$Z\to \mu^+ \mu^-$ decay. 
The impact of the Higgs signal at 125~GeV is not shown.
The ILC projections are compared with the 
observed limit at LEP from the $\phi \to b \bar b$
searches~\cite{Barate:2003sz}.
    The solid horizontal lines denote the indirect sensitivities at the HL-LHC 
and the ILC with $\sqrt{s}=250$~GeV
from the projection for the accuracies
    on the couplings of the observed Higgs boson at 125~GeV
   \cite{deBlas:2019rxi} (see text). \label{fig:CombinedLimits}
    }
  \end{figure}

In Fig.~\ref{fig:CombinedLimits} we show our results for the 
$S_{95}$ projections at the ILC 
for the analysis making use of the decay $\phi \to b \bar b$ decay and
and for the recoil method, where only the information from 
$Z\to \mu^+ \mu^-$ has
been used~\cite{georg-mext}
(the expected limits do not take into account the impact 
of the Higgs signal at 125~GeV).
These results are compared with the observed limit from the 
$\phi \to b \bar b$ search at 
LEP~\cite{Barate:2003sz}.
As one can see from Fig.~\ref{fig:CombinedLimits},
the projected ILC limits yield an improvement of more than a factor of 10
in $S_{95}$ for the mass range between about 60~GeV and 100~GeV
compared to the existing limits from LEP already for the 
moderate luminosity of ${\cal L}=500$~fb$^{-1}$ at
$\sqrt{s}=250$~GeV.%
\footnote{For one of the two types of analyses addressed in our
paper, namely the analysis using the recoil method, meanwhile 
an ILC study has been performed
with full ILD detector simulation~\cite{Wang:2018awp,Wang:2019mzd}. 
The results obtained in 
Ref.~\cite{Wang:2018awp,Wang:2019mzd} 
are in good qualitative agreement \htGWb{(in view of the more complete
treatment of backgrounds arising from events involving photons)}
with the curve in  
Fig.~\ref{fig:CombinedLimits} showing our projection for the ILC based on the
recoil method.}

For comparison, in Fig.~\ref{fig:CombinedLimits} also 
the indirect sensitivity at the HL-LHC from the projection for the accuracies
on the couplings of the observed Higgs boson at 125~GeV~\cite{deBlas:2019rxi} 
is indicated. It corresponds to the area of the plot above the 
solid horizontal black line labelled as ``HL-LHC''. 
The rate measurements of the state at 125~GeV
provide an indirect sensitivity to the squared coupling 
$(g_{\phi ZZ})^2$
of an additional light (or heavy) Higgs boson $\phi$ under
the assumption that the sum rule of Eq.~\eqref{eq:g2} is valid. The projected
accuracy on the coupling of the Higgs boson at 125~GeV, $h(125)$, to
$Z$~bosons, $\Delta g_{h(125)ZZ}$, translates under this assumption into an
upper bound on the squared coupling of the light Higgs boson according to
\begin{equation}
\frac{(g_{\phi ZZ})^2}{(g^{\rm SM}_{HZZ})^2} \leq
1 - \left(1 - \left|\Delta g_{h(125)ZZ}\right|\right)^2
\end{equation}
up to higher-order corrections to the sum rule.
The indirect sensitivity achievable at the HL-LHC that is indicated
in Fig.~\ref{fig:CombinedLimits} is
based on the
projection for the $2 \sigma$ accuracy of the $g_{h(125)ZZ}$ coupling
from Tab.~4 of 
Ref.~\cite{deBlas:2019rxi}, where an integrated luminosity of 3000~fb$^{-1}$
was assumed. While the indirect sensitivity at the HL-LHC exceeds the
observed limit from LEP for $m_\phi \gsim 85$~GeV, 
Fig.~\ref{fig:CombinedLimits} clearly demonstrates that the sensitivity of the
direct search at the ILC will much surpass the indirect sensitivity at the
HL-LHC even for the ultimate accuracy reachable at the HL-LHC.

The indirect sensitivity of the ILC with
$\sqrt{s}=250$~GeV from the rate measurements of the state at 125~GeV,
where as for the HL-LHC we have used the
projections listed in Tab.~4 of Ref.~\cite{deBlas:2019rxi}, 
is indicated in Fig.~\ref{fig:CombinedLimits} as the area above the
solid horizontal black line that is labelled as ``ILC''. 
Fig.~\ref{fig:CombinedLimits} shows that this
indirect
sensitivity at the ILC would be similar to the direct reach via the recoil
method, which could provide important complementary information for 
determining the nature of a possible excess in the direct searches for
additional light Higgs bosons at the ILC. It should be noted
that the sensitivity of the direct
ILC searches for additional Higgs bosons making use of the 
$\phi \to b \bar b$ decay mode will significantly improve even on the
indirect sensitivity of the ILC based on its high-precision measurement of
the coupling of $h(125)$ to $Z$ bosons.

\section{Conclusions\label{sec:5}}

In this paper we have pointed out that the ILC at $\sqrt{s}=250$~GeV has a 
large physics potential in the direct search for additional light Higgs
bosons. Such a light Higgs boson would be expected to have a heavily
suppressed coupling to gauge bosons as compared to a SM-like Higgs boson at
the same mass, which could therefore be below the existing limits from LEP.
Via those direct searches
the ILC would probe favoured parameter regions of various extensions of the
SM, which have recently received considerable attention also in view of the
excess over the background expectation that has been reported for the 
$\phi \to \gamma\gamma$ searches at CMS in the vicinity of a
long-standing excess in the LEP Higgs searches. 

We have performed a generator-level study for searches via the 
Higgs-strahlung process at a future $e^+e^-$ collider for both the analysis
type using the $\phi \to b \bar b$ decay and for the decay-mode independent
search via the recoil method. In a first step we have validated our approach
with the existing results for the search in the 
$\phi \to b \bar b$ channel, from the LEP combination, and for the search
utilising the recoil method, from the OPAL collaboration. We 
determined normalisation factors for the effective luminosities that we
employ to approximately account for signal efficiencies and
detector effects. We demonstrated that with this input our method reproduces
the expected limits from LEP and OPAL for the two types of analyses very
well. 

After this validation we applied our method for deriving the expected
limits for the ILC at $\sqrt{s}=250$~GeV for both types of analyses. We used
the normalisation factors $c_{\rm bb}$ and $c_{\rm recoil}$ that we 
determined for LEP~2
but assumed as ILC conditions a centre-of-mass energy of 250~GeV, 
beam polarisation of
$P_{e^-}=-80\%$ for the electron beam and $P_{e^+}=+30\%$ for the positron
beam and used as a very conservative approach the rather modest luminosity
of 500~fb$^{-1}$.
Our results show that the ILC at $\sqrt{s}=250$~GeV 
will improve the LEP limits 
in  the sensitivity to a light Higgs boson with reduced couplings to gauge bosons 
for the most interesting mass range between about 60~GeV and 100~GeV
by more than an order of magnitude. This sensitivity of the direct search for
additional light Higgs bosons at the ILC will go much beyond the indirect 
sensitivity of the HL-LHC from the
rate measurements of the detected state at 125~GeV 
even for the projected ultimate accuracy reachable at the HL-LHC with 
3000~fb$^{-1}$. It is interesting to note in this context that the indirect
sensitivity of the ILC with
$\sqrt{s}=250$~GeV from the rate measurements of the state at 125~GeV 
is similar to the direct search reach of the ILC via the recoil method, while
the direct searches for additional light Higgs bosons at the ILC utilising 
the $\phi \to b \bar b$ decay will significantly improve even on the ultimate
indirect ILC sensitivity.

The physics potential for the direct searches at the ILC with
$\sqrt{s}=250$~GeV discussed in this paper complements and significantly
enhances the ILC physics progmamme for precision measurements in the Higgs
and the electroweak sector.

\vspace{-1mm}
\section*{Acknowledgements}
We thank P.~Bechtle, K.~Desch, S.~Heinemeyer, F.~Lika, J.~List and A.~Raspereza for useful discussions.
G.M.P.\ and G.W.\ acknowledge support
by the Deutsche Forschungsgemeinschaft (DFG, German Research
Foundation) under Germany‘s Excellence Strategy -- EXC 2121 ``Quantum
Universe'' -- 390833306.

\pagestyle{plain}             

\end{document}